\documentclass[twocolumn,superscriptaddress,prapplied]{revtex4-1}
\usepackage{graphicx}
\usepackage{amssymb}
\usepackage{amsmath}
\usepackage{epsfig}
\usepackage{color}
\usepackage{mathtools}
\usepackage[colorlinks,linkcolor=blue,anchorcolor=blue,citecolor=blue,urlcolor=blue]{hyperref}

\begin{document}

\title{Near-infrared-enhanced charge state conversion for low power optical nanoscopy with nitrogen vacancy center in diamond}
\author{Xiang-Dong Chen}
\author{Shen Li}
\author{Ao Shen}
\author{Yang Dong}
\author{Chun-Hua Dong}
\author{Guang-Can Guo}
\author{Fang-Wen Sun}
\email{fwsun@ustc.edu.cn}

\affiliation{Key Lab of Quantum Information, Chinese Academy of Sciences, School of physics, University of Science and Technology of China, Hefei, 230026, P.R. China}
\affiliation{Synergetic Innovation Center of Quantum Information $\&$ Quantum Physics, University of Science
and Technology of China, Hefei, 230026, P.R. China}
\date{\today}

\begin{abstract}
The near-infrared (NIR) optical pumped photophysics of nitrogen vacancy (NV) center in diamond was experimentally studied by considering both the charge state conversion and stimulated emission. We found that the NIR laser can help to highly enhance the charge state conversion rate, which can be applied to improve the performance of charge state depletion nanoscopy. Using a doughnut-shaped visible laser beam and a Gaussian-shaped NIR laser beam for charge state manipulation, we developed a low power charge state depletion nanoscopy for NV center. A spatial resolution of 14 nm was achieved with the depletion laser intensity approximately three orders lower than that used for the stimulated emission depletion nanoscopy with NV center. With high spatial resolution and low laser power, the nanoscopy can be used for nanoscale quantum sensing with NV center. And our study on the charge state conversion can help to further optimize the NV center spin state initialization and detection.
\end{abstract}
\maketitle

\section{Introduction}
Quantum sensing with high spatial resolution is a useful tool for nanoscience. As one of the most promising quantum sensors, nitrogen vacancy (NV) center in diamond has proved its capability for the temperature and electromagnetic field sensing with high spatial resolution and sensitivity \cite{Jacques-science2014-domainwall,Meriles-nc2015-tem,Walsworth2016magsensing,lukin-nature-therm}.
High density NV center ensemble has been applied for the quantum imaging based on optical far-field microscopy \cite{holl-2012-screp,roch-2015epjd,holle2015magimaging,Awschalom-apl2010-magnetic}. Moreover, due to the noninvasive property, NV centers with optical far-field microscopy are especially suitable for labeling and sensing in living cells \cite{lukin-2013naturecell,Treussart2009acs,changhc-2016acr,Hollenberg-nn-2011}.

The spatial resolution of traditional optical far-field microscopy is generally limited by the optical diffraction. Recently, several different types of super-resolution optical far-field microscopies have been developed to locate NV center with the resolution below diffraction limit \cite{adma2012sil,hell-2009,acsnano-2013-sted,chang-2011sted,hell-nl-micro,chen201501,lukin-farfield,hell-2009epl,cui-prl,pengxi-2014-rcs,nl-2013-mw,wra2014PNASstorm}. Among them, the reversible saturable optical linear fluorescence transitions (RESOLFT) \cite{adma2012sil,hell-2009,hell-2009epl,acsnano-2013-sted,chang-2011sted,hell-nl-micro,chen201501,lukin-farfield} has the capability to highly improve the spatial resolution of quantum sensor with NV center. One of the most typical RESOLFT type methods is stimulated emission depletion (STED) nanoscopy \cite{adma2012sil,hell-2009,acsnano-2013-sted,chang-2011sted}. With high power doughnut-shaped depletion laser, STED nanoscopy can image the NV center with nanoscale spatial resolution. It has also been successfully applied for sub-diffraction resolution electron spin resonance measurement \cite{hell-2011}.

However, high laser power would induce heating of NV center \cite{chang-2015nl-tem,heating2016}, and subsequently decrease the fidelity of spin state manipulation \cite{temshift} and the sensitivity of quantum sensor.  And for biological cell imaging, high laser power would cause significant photondamage \cite{photodamge2015,hell2011acie}. It is crucial to drop the pump intensity for the RESOLFT type super-resolution microscopy. Recently, based on the photon induced charge state conversion in NV center, a low pump power charge state depletion (CSD) nanoscopy has been developed \cite{hell-nl-micro,chen201501}. Using a doughnut-shaped visible laser beam for charge state depletion, super-resolution optical imaging of NV center is obtained. High spatial resolution local field sensing with NV center has been realized through the CSD nanoscopy \cite{chen201601}. In this work, we used near-infrared (NIR) laser additionally on the visible laser pump to much enhance the transition rate between the charge states and highly decrease the power of laser for the CSD nanoscopy.

The charge state conversion between the negatively charged NV$^{-}$ and the neutrally charged NV$^{0}$ is the base of CSD nanoscopy \cite{hell-nl-micro,chen201501}. It has also been applied for the NV$^{-}$ spin qubit initialization and detection \cite{wra-lowtem-2012,chen20152prb,lukin2014spincharge}, which enables the NV center for quantum information and metrology \cite{wra2006review,elestr-prb2006}. The zero phonon lines (ZPLs) of NV$^{0}$ and NV$^{-}$ are 575 and 637 nm, respectively. Therefore, the fluorescence of NV center can be switched during the charge state conversion. Usually the charge state of NV center is controlled with visible light pumping \cite{excwave-prl-2012,ioni-arxiv-2012,wra-lowtem-2012,chen20132,Mizuochi-2016prb}. Specially, 532 nm green laser has been used for NV$^{-}$ initialization with a probability about 75$\%$, and 637 nm red laser has been used for NV$^{0}$ initialization with 95$\%$ probability \cite{chen20132,chen20152prb}. In this way, the 532 and 637 nm lasers are optimally used as depletion laser in CSD nanoscopy \cite{hell-nl-micro,chen201501} with nanoscale resolution. In order to further decrease the laser power of CSD nanoscopy without dropping the imaging resolution, it is highly required that the transition rate should be primarily increased with the depletion laser.

Here, the photophysics of NV center with NIR laser excitation was studied in detail by considering the photon induced charge state conversion and stimulated emission. We found that the NIR laser can help to highly enhance the charge state conversion rate. Then, a two-depletion-laser pumped CSD nanoscopy was developed by using doughnut-shaped 532 nm laser beam and an additional Gaussian-shaped 780 nm laser beam as depletion lasers. Without sacrificing the image resolution, the power of visible depletion laser is decreased approximately 10 times by applying a 0.4 mW 780 nm laser beam. The depletion laser intensity of this optimized CSD nanoscopy is three orders lower than that used in STED nanoscopy with the similar resolution. This technique can be used for nanoscale quantum sensor with NV center. Specially for the imaging of biological cell, the decrease of power and replacing visible laser with NIR depletion laser can reduce the photodamage induced by laser pumping of CSD nanoscopy \cite{photodamge2015,hell2011acie}.

\section{Photophysics of NV center with NIR laser excitation}
\begin{figure}
  \centering
  \includegraphics[width=7.5cm]{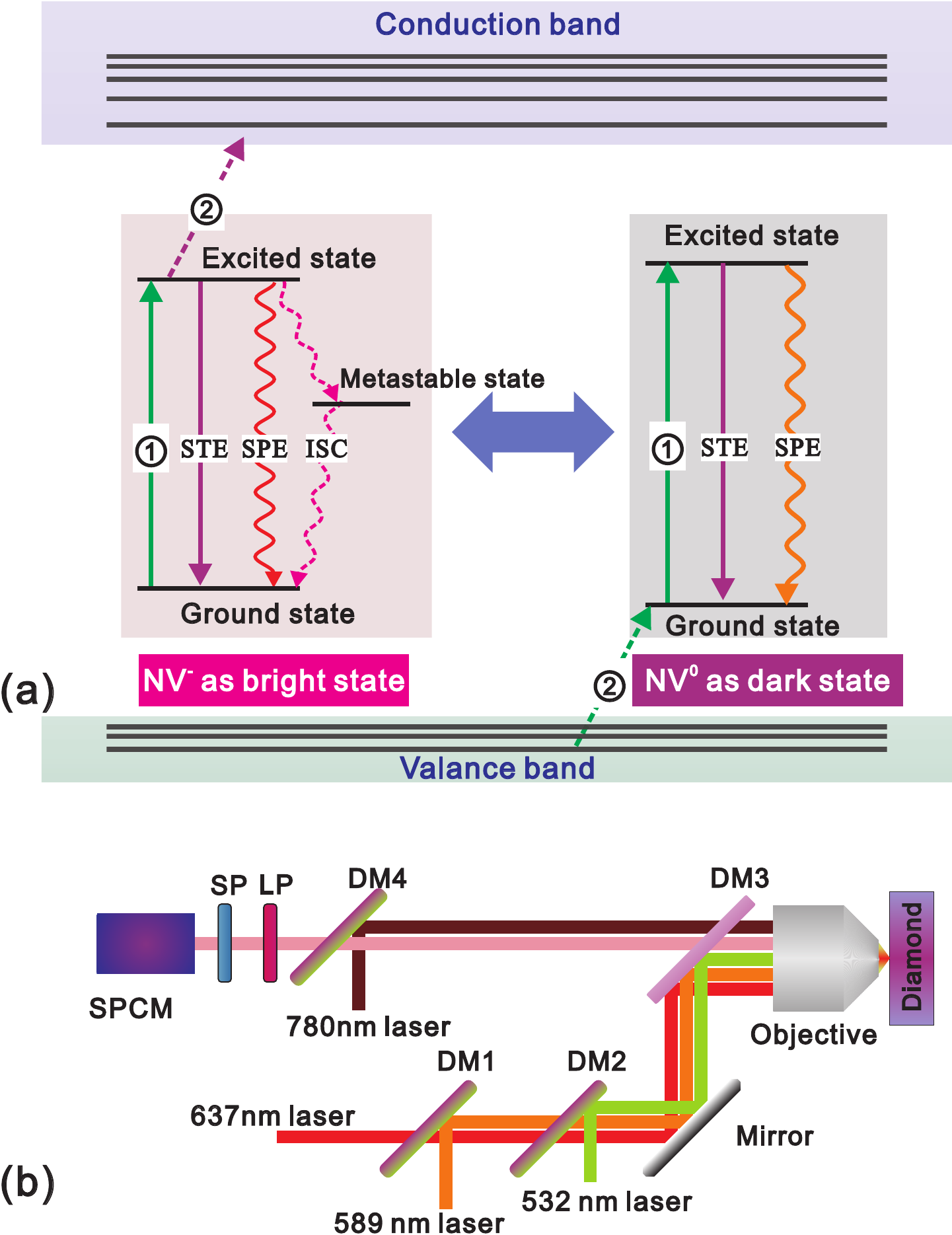}\\
  \caption{(a) The level scheme for NV center charge state conversion. The transitions of electron during two-photon charge state conversion processes are indicated as $\textcircled{1}$ and $\textcircled{2}$ for the first and second transitions, respectively. STE, stimulated emission; SPE, spontaneous emission; ISC, intersystem crossing. (b) Diagram of the experimental setup. DM1-3 are long-pass dichroic mirrors with edge wavelengths of 605, 536.8 and 658.8 nm, respectively. DM4 is short-pass dichroic mirror with an edge wavelength of 746 nm. LP, long-pass filter with an edge wavelength of 668.9 nm; SP, short-pass filter with an edge wavelength of 761 nm.}\label{figlevel}
\end{figure}

The charge state conversion of NV center pumped by visible wavelength photons has been proved to be a two-photon process \cite{ioni-arxiv-2012,wra-lowtem-2012,chen20132}. As shown in Fig. \ref{figlevel}(a), a photon firstly pumps an NV center to the excited state of NV$^{-}$ or NV$^{0}$ (indicated as $\textcircled{\footnotesize{1}}$). The photon energy of this transition should match the energy gap between excited state and ground state. Then, the NV center absorbs another photon (indicated as $\textcircled{\footnotesize{2}}$) to release an electron to the conduction band (ionization process, NV$^{-}$ to NV$^{0}$ conversion) or capture an extra electron from the valance band (recombination process, NV$^{0}$ to NV$^{-}$ conversion). The second process is less spectral sensitive than the first process, as it can be seen as the transition between bound state and continuum states. Usually, both the ionization and recombination processes can be pumped by the photons with same wavelength. The steady state population of NV$^{-}$ or NV$^{0}$ is determined by the ratio of recombination rate to ionization rate. To further optimize the NV center charge state manipulation and detection for CSD nanoscopy, it is important to fully understand the optical dynamics during the charge state conversion process.

Recently the charge state conversion \cite{Dutt-prb2016charge,Bassett2016csc,gumin2013oe} and stimulated emission \cite{Greentree2016ste,hell-2011} have been studied in NIR spectral range. These transitions would compete with each other, and extend the method for NV center optical manipulation. Though the mechanism is still not well understood, high signal to noise ratio spin-readout protocol has already been proposed based on NIR photon excitation \cite{Bassett2016csc}. Applying NIR laser for charge state depletion, we expect that the performance of CSD nanoscopy can be further improved.

\subsection{Experimental setup}

The sample was a CVD diamond plate (Element Six). NV centers were produced by 20 KeV nitrogen ion implanting. In the experiment, we used a 780 nm laser from Ti:sapphire laser system to study the NIR photon pumped photophysics. In addition, 532 and 637 nm diode lasers were used for optical manipulation of NV center. A weak 589 nm laser was used for the charge state detection. The experimental setup is shown in Fig.\ref{figlevel}(b). The laser pulses were controlled with acousto-optic modulators and combined by dichroic mirrors (DMs). The laser beams were then focused on the diamond sample through an objective with 0.95 numerical aperture. Since the photons of NV$^{-}$ and NV$^{0}$ are spectral separated, only the phonon sideband of NV$^{-}$ charge state was detected with a single-photon-counting-module (SPCM), while the fluorescence of NV$^{0}$ was blocked by a long pass optical filter. And the 780 nm laser and stimulated emission of NV center were blocked by short pass filter. Therefore, the detected fluorescence intensity increases linearly with the population of NV$^{-}$.
\subsection{Dynamics of charge state conversion and stimulated emission}

\begin{figure}
  \centering
  \includegraphics[width=8.5cm]{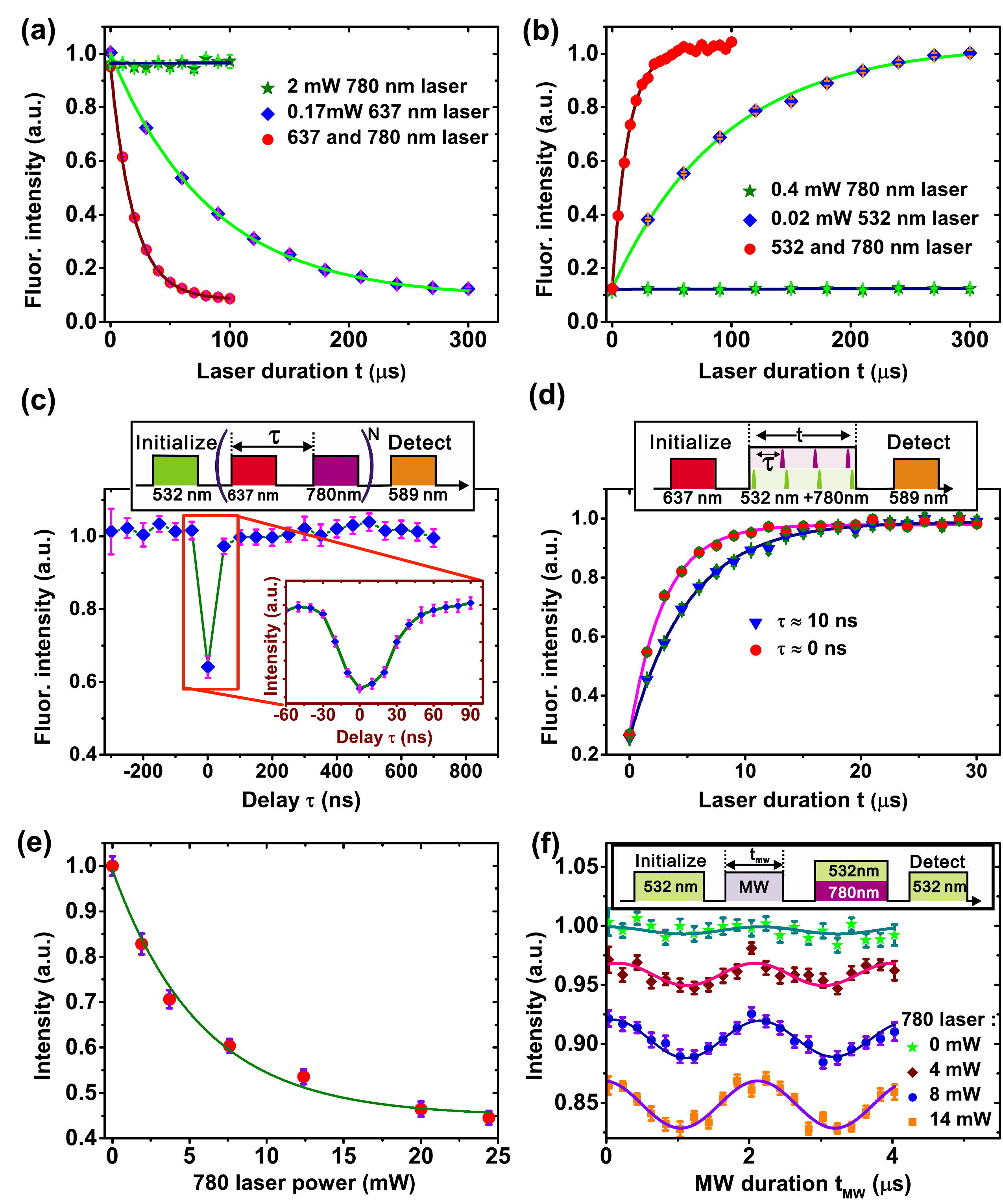}\\
  \caption{ (a)(b) The charge state conversion with visible and NIR lasers excitation. The NV center is firstly initialized by 532 (a) or 637 (b) nm laser, and then pumped by visible and NIR laser. Finally, the charge state is detected by pumping the NV center with 589 nm laser. (c) The charge state conversion pumped by 780 and 637 nm laser pulses with different delays. The width of 780 and 637 nm laser pulses is 30 ns, and the pulses repeat $N=400$ cycles. Average power: 637 nm laser, 0.5 mW; 780 nm laser, 5 mW. (d) The charge state conversion process pumped by 780 and 532 nm pulsed laser. Average power: 532 nm laser, 0.06 mW; 780 nm laser, 5 mW. (e) The fluorescence intensity of NV center under coincident excitation with a fixed 532 nm laser and a varied 780 nm laser. (f) The spin state Rabi oscilation of NV$^{-}$ measured after 532 and 780 nm laser excitation. The power of 780 nm laser is indicated in the figure. The 532 nm laser power is fixed at 0.65 mW. The durations of 532 and 780 nm laser are 300 ns.}\label{figtransition}
\end{figure}

By detecting the fluorescence of NV center, the charge state conversion pumped by 532, 637 and 780 nm lasers was measured, as shown in Fig. \ref{figtransition}(a)(b). Since the ZPL of NV center is at 637 nm for NV$^{-}$ and 575 nm for NV$^{0}$, the energy of 780 nm photon is not high enough to pump the NV center from the ground state to excited state. Therefore, no signal of charge state conversion is detected when the NV center is pumped by a single 780 nm laser. However, the charge state conversion pumped by 532 or 637 nm laser can be significantly accelerated by applying an additional 780 nm laser beam. It indicates the second transition of the two-photon charge state conversion can be pumped by 780 nm laser.

For the NIR laser assisted the second transition of ionization process, it has been suggested that the electron could be pumped to the conduction band from both the excited state and metastable spin singlet state of NV$^{-}$ \cite{Bassett2016csc}. The lifetime of NV$^{-}$ excited state is approximately 12 ns, while the lifetime of metastable state is approximately 200 ns. To analyze the exact optical dynamics under 780 nm laser excitation, we measured the charge state conversion with various time delays between 780 nm laser pulse and 637 nm laser pulse in Fig. \ref{figtransition}(c). The 637 nm laser firstly pumps the NV$^{-}$ from ground state to excited state. And then the 780 nm laser can pump the second transition of ionization process. The results show that only when the two laser pulses simultaneously pump the NV center, the charge state conversion can be accelerated by 780 nm laser. The time interval between the first and the second transitions of ionization is much shorter than 200 ns. It proves that the second transition of ionization process pumped by 780 nm laser is mainly the transition of electron from excited state of NV$^{-}$ to the conduction band of diamond. The transition from the metastable state can be neglected.

In order to maximize the accelerating effect of 780 nm laser, we use pulsed visible laser and 780 nm laser for charge state conversion, and synchronize the two laser pulses. The charge state conversions with different relative delays between 532 and 780 nm pulsed laser (76 MHz repetition rate) are shown in Fig. \ref{figtransition}(d). The conversion rates are approximately 0.19 $\mu s^{-1}$ with 10 ns delay and 0.34 $\mu s^{-1}$ with 0 ns delay in the results.

The stimulated emission can also be pumped by NIR laser \cite{Greentree2016ste,hell-2011}. It depletes the excited state of NV center and competes with the two-photon charge state conversion. Subsequently, the stimulated emission would decrease the charge state conversion rate. Therefore, we have to verify the effect of stimulated emission pumped by 780 nm laser.
As shown in Fig. \ref{figtransition}(e), the measured spontaneous emission intensity decreases with the 780 nm laser power due to stimulated emission. This effect has been used for the STED nanoscopy of NV center \cite{hell-2009}. And a more persuasive evidence of stimulated emission is the spin state population preservation under 780 nm laser pumping \cite{hell-2011}. As shown in Fig. \ref{figtransition}(f), the Rabi oscillation of NV$^{-}$ electron spin state is pumped by resonant microwave pulse, then the spin state is detected after 780 and 532 nm laser excitation with a duration of 300 ns. The spin-selective intersystem crossing (ISC) process induced by 532 nm laser excitation polarizes the electron spin state of NV$^{-}$ \cite{wra2006review,elestr-prb2006}, and subsequently disturbs the results of Rabi oscillation. Therefore, with single 532 nm laser pumping, the Rabi oscillation of NV$^{-}$ can not be detected. In contrast, with simultaneous 532 and 780 nm lasers excitation, the Rabi oscillation of NV$^{-}$ can be detected. The visibility of Rabi oscillation increases with the power of 780 nm laser. It indicates that the spin-conserving stimulated emission is pumped by 780 nm laser. As the charge state conversion has been proved to be a spin-depolarization process \cite{chen20152prb}, the results also prove that the stimulated emission is faster than the charge state conversion pumped by 780 nm laser.

\begin{figure}
  \centering
  \includegraphics[width=8.5cm]{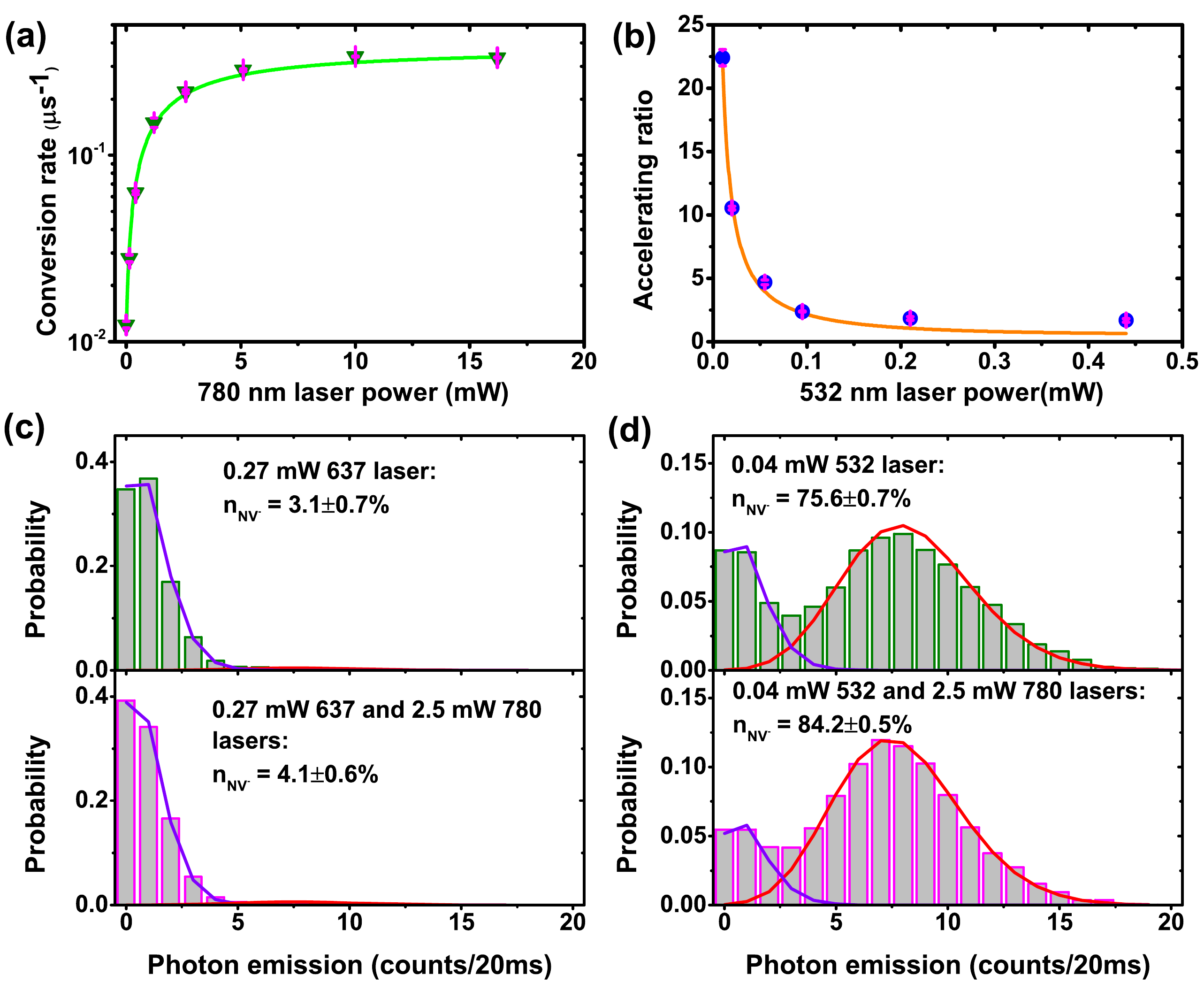}\\
  \caption{(a) The charge state conversion rates with 0.02 mW 532 nm laser and different 780 nm lasers. (b) The 0.8 mW 780 nm acceleration effect with different 532 nm laser power. (c)(d) The steady state population of charge state pumped by different lasers. The histograms of photon counts of a single NV center was measured by the single-shot readout charge state method. High photon counting rate presents the NV$^{-}$ charge state.}\label{figcscrate}
\end{figure}

\subsection{Power dependent conversion rate}

According to the above results, the charge state conversion is affected by 780 nm laser through both stimulated emission and the second transition of charge state conversion. Using the two-photon charge state conversion model \cite{wra-lowtem-2012,ioni-arxiv-2012}, the ionization(recombination) rate can be simply written as:$\frac{\gamma_{ge}\gamma_{bc}}{\gamma_{eg}+\gamma_{ge}+\gamma_{bc}}$ (See Appendix for detals).
$\gamma_{ge}$ is the transition rate from ground state to the excited state (the first transition of charge state conversion), which is pumped by visible light beam in our experiment. $\gamma_{eg}$ denotes the decay from excited state to the ground state. It includes stimulated emission, spontaneous emission and ISC through metastable state (only for ionization process): $\gamma_{eg}=\gamma_{STE}+\gamma_{SPE}+\gamma_{ISC}$. $\gamma_{bc}$ is the transition between bound state and continuum states (the second transition of charge state conversion), which can be pumped by both visible and NIR photons: $\gamma_{bc}=\gamma_{bc,532}+\gamma_{bc,780}$.

The results in Fig. \ref{figtransition}(a)(b) show that the charge state conversion rate is increased by applying 780 nm laser. It indicates that $\frac{\gamma_{bc,780}}{\gamma_{bc,532}}>\frac{\gamma_{STE}}{\gamma_{ge}+\gamma_{SPE}+\gamma_{ISC}}$ in our experiments. Therefore, it can be derived that the charge state conversion rate would be increased by increasing both visible and NIR laser power. However, the upper limit of conversion rate is still determined by the visible laser power. In Fig. \ref{figcscrate}(a)(b), we show the charge state conversion rates with simultaneously 532 and 780 nm laser pumping. For the charge state conversion pumped by 0.02 mW 532 nm laser, the total conversion rate can be improved 27 times with high 780 nm laser power. And the NIR laser accelerating ratio, defined as the ratio of conversion rate with 780 nm laser to that without 780 nm laser, decreases with 532 nm laser power.

Though fast stimulated emission and charge state conversion both have been excited by 780 nm laser, there is only a small change in the steady state population. In Fig.\ref{figcscrate} (c)(d), the steady state populations of NV center charge state are measured with single-shot charge state readout method \cite{ioni-arxiv-2012,chen20132}. High NV$^{0}$ population is obtained by 637 and 780 nm lasers excitation, and high NV$^{-}$ population is obtained by 532 and 780 nm lasers excitation. As the steady state population is determined by both recombination and ionization, the results in Fig. \ref{figcscrate}(d) demonstrate that the ratio of recombination rate to ionization rate increases only 1.6 times by applying 2.5 mW 780 nm laser. In contrast, the total charge state conversion rate increases approximately 10 times with similar 780 nm laser power. It indicates that the ionization and recombination processes are both accelerated by applying 780 nm laser. Therefore, the 780 nm NIR photon can be seen as a purely accelerator for charge state conversion, and then can be applied for CSD nanoscopy.

\section{CSD nanoscopy with NIR depletion laser}

For the CSD nanoscopy, the depletion laser should have the capability for fast and high fidelity charge state manipulation. Usually, 532 nm laser or 637 nm laser are used, as they can initialize the charge state to NV$^{-}$ or NV$^{0}$ with high probability. The conversion rate with 532 nm laser excitation has been proved to be higher than that with same power 637 nm laser \cite{chen201501}. Here, we use 532 nm laser as the charge state depletion laser for CSD nanoscopy.

\begin{figure}
  \centering
  \includegraphics[width=8.0cm]{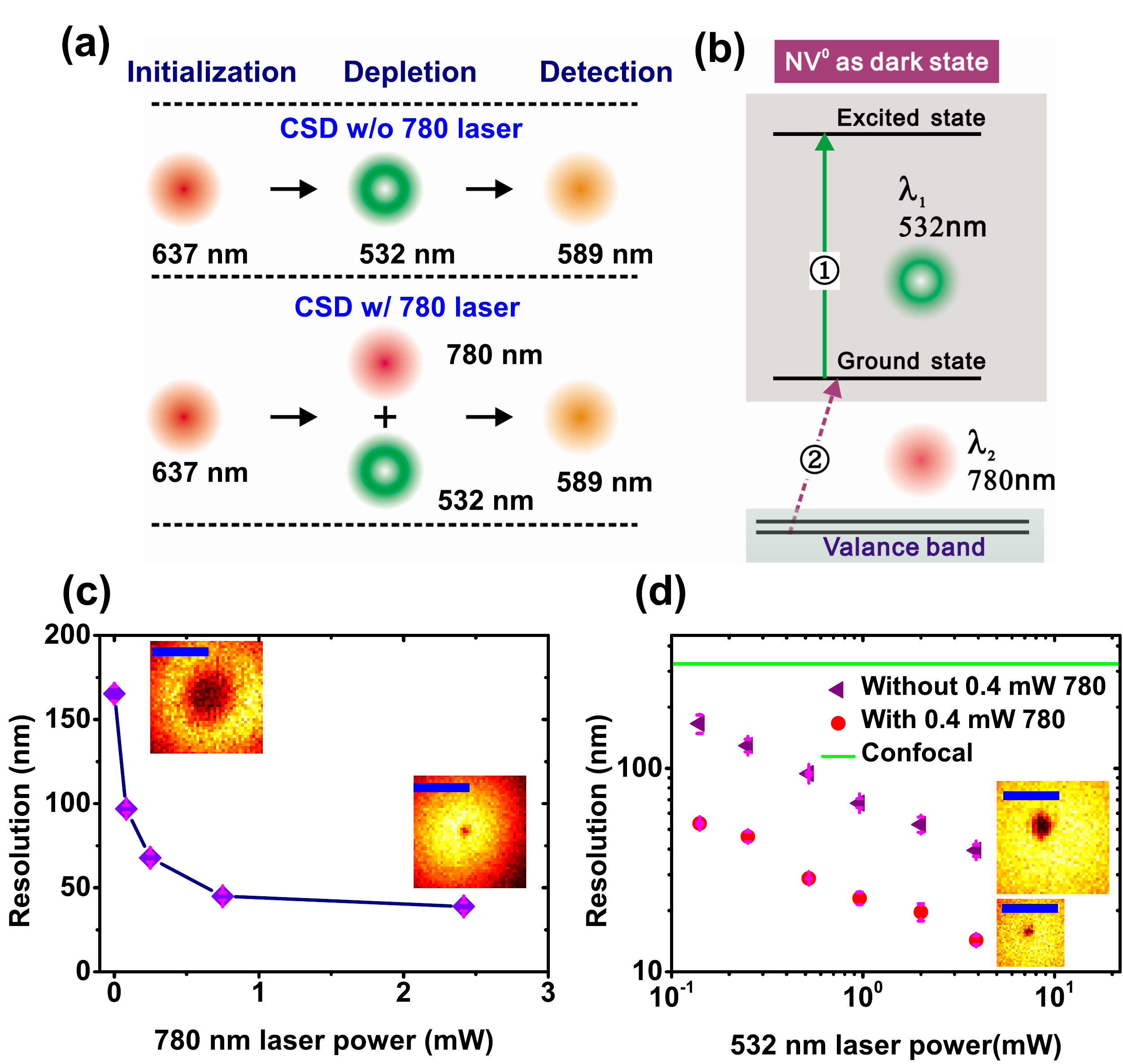}\\
  \caption{(a) The laser pulse sequence for single-laser and two-laser pumping CSD nanoscopy. (b) The schematic of level transition for CSD nanoscopy. (c)(d) The CSD nanoscopy resolution with different 780 and 532 nm laser pumping. 532 nm laser in (c), 0.14 mW. The duration of depletion laser is 200 $\mu s$. The scale bars, 200 nm (c) and 100 nm (d).}\label{figresolution}
\end{figure}

The laser pulse sequence is shown in Fig.\ref{figresolution}(a). A Gaussian-shaped 637 nm laser pulse is firstly applied to initialize the NV center to NV$^{0}$ charge state. Then a doughnut-shaped 532 nm laser beam, which is generated by vortex phase mask, is used to pump NV$^{0}$ to NV$^{-}$ (recombination process) with laser intensity dependent transition rate. Only the NV centers at the 532 nm laser beam center maintain NV$^{0}$ charge state. At last, a Gaussian-shaped 589 nm laser beam is used for NV center charge state detection. With long pass filters, only the fluorescence of NV$^{-}$ is detected for CSD nanoscopy. Therefore, the NV center is presented as a dark point in the images. As the charge state conversion rate increases with the power of visible depletion laser, the spatial resolution of CSD nanoscopy would be improved by increasing the power of 532 nm doughnut-shaped laser beam.

Here, instead of increasing the doughnut-shaped laser beam power, we apply an additional Gaussian-shaped 780 nm laser beam to increasing the charge state conversion rate. The laser sequence is shown in the downside of Fig.\ref{figresolution}(a). The charge state depletion is pumped by 532 and 780 nm lasers. As the single 780 nm laser beam can not pump charge state conversion, the total effect of 532 and 780 nm lasers to the charge state conversion will maintain the doughnut shape. And the charge state conversion accelerating the effect of 780 nm laser is maximum at around the center of 532 nm doughnut-shaped beam. In this way, the recombination process can be seen as the transitions in Fig.\ref{figresolution}(b). The ground state to excited state transition is pumped by 532 nm doughnut-shaped laser beam, and the transition between bound state and continuum states is mainly pumped by 780 nm Gaussian-shaped laser beam.

The resolution of the two-depletion-laser CSD nanoscopy is determined by both 532 and 780 nm lasers' powers. As shown in Fig. \ref{figresolution}(c)(d), the resolution is improved 4 times by increasing the 780 nm laser power. However, 780 nm laser can not infinitely improve the resolution. It is because the charge state depletion rate is limited by the ground state to excited state transition rate $\gamma_{ge}$, which is determined by the 532 nm  laser power. With 0.4 mW 780 nm laser beam, the resolution has already been improved 3 times. In contrast, the 532 nm doughnut beam power is needed to increase approximately 10 times to obtain the similar resolution without 780 nm laser beam. Specifically, with 3.9 mW 532 nm doughnut-shaped laser beam and 0.4 mW 780 nm Gaussian-shaped laser beam, a resolution of 14 nm is reached.  The total intensity of 532 and 780 nm depletion laser is estimated to be 1.2 MW/cm$^{2}$. In comparison, the intensity of several GW/cm$^{2}$ is required for the similar resolution of STED \cite{adma2012sil,hell-2009,acsnano-2013-sted}.

\section{Summary}
Recent experiments show that the stimulated emission of NV center changes with the NIR laser wavelength \cite{Greentree2016ste}. Low stimulated emission rate could improve the charge state conversion rate pumped by NIR laser. Therefore, the accelerating effect for charge state conversion could be optimized by changing the NIR photon wavelength. And the CSD nanoscopy laser power can be further decreased.

In conclusion, we studied the optical dynamics of NV center pumped by NIR laser. Both charge state conversion and stimulated emission are observed with 780 nm laser pump. The charge state conversion process pumped by visible light is significantly accelerate by applying an additional 780 nm laser. Subsequently, we use 532 nm doughnut-shaped beam and 780 nm Gaussian-shaped beam as depletion lasers for CSD nanoscopy. With an additional 0.4 mW 780 nm laser beam, the total power of depletion laser can be decreased approximate 10 times for a fixed resolution. The decrease of depletion laser power can reduce the heating and photodamage to sample. Combining the CSD nanoscopy and quantum sensing, nanometer scale quantum imaging can be realized with NV center ensemble.

\section*{Acknowledgment}
This work was supported by the Strategic Priority Research Program(B) of the
Chinese Academy of Sciences (Grant No. XDB01030200), the National Natural
Science Foundation of China (Nos. 11374290, 91536219, 61522508, 11504363),
the Fundamental Research Funds for the Central Universities, the China
Postdoctoral Science Foundation (No.2016T90565), and the Foundation for the
Author of National Excellent Doctoral Dissertation of China.

\appendix*
\section{The model of conversion rate}
\label{app:model}

\begin{figure}
  \centering
  \includegraphics[width=4.0cm]{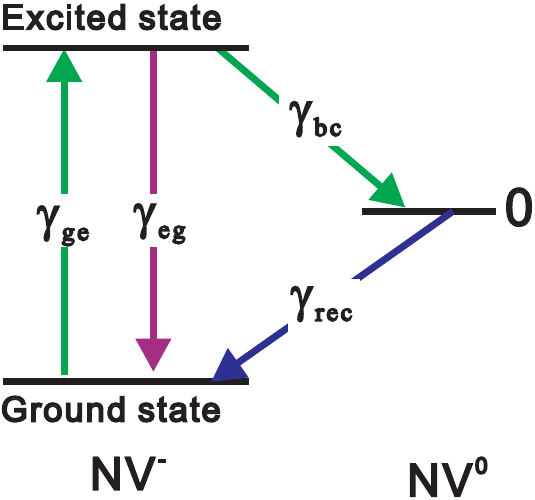}\\
  \caption{ Illustrations of energy levels and transition rates for ionization rate analysis.}\label{figap}
\end{figure}

During the charge state conversion, the population of NV center charge state evolves as a exponential function. The evolution rate is $\gamma=\gamma_{ion}+\gamma_{rec}$, where $\gamma_{ion}$($\gamma_{rec}$) is the ionization (recombination) rate \cite{ioni-arxiv-2012,chen201501}. To analyze the form of ionization (recombination) rate, a three-level model is used, as in Fig.\ref{figap}. It includes the ground state (G) and excited state (E) of NV$^{-}$, and an effective state (0) of NV$^{0}$. Since the ionization and recombination have the similar dynamics, we focus on the ionization process here. The rate equations are:
\begin{eqnarray}
  \dot{n_{G}} &=& -\gamma_{ge}n_{G}+\gamma_{eg}n_{E}+\gamma_{rec}n_{0}, \\
  \dot{n_{E}} &=& \gamma_{ge}n_{G}-\gamma_{eg}n_{E}-\gamma_{bc}n_{E}, \\
  \dot{n_{0}} &=& \gamma_{bc}n_{E}- \gamma_{rec}n_{0},
\end{eqnarray}
where $n_{i}$ denotes the population of $i$ state. $\gamma_{ge}$ ($\gamma_{eg}$) is the the transition rate from ground state to excited state (excited state to ground state), and $\gamma_{bc}$ is the transition rate from NV$^{-}$ excited state to NV$^{0}$ (transition between bound state and continuum states). The recombination rate is presented by $\gamma_{rec}$.

In order to obtain the ionization rate, we set the recombination rate $\gamma_{rec}$ to zero and solve the rate equations. The solution shows that the effective ionization rate is
\begin{equation}\label{ionrate1}
  \gamma_{ion}=\frac{1}{2}(\gamma_{all}-\sqrt{\gamma_{all}^{2}-4\gamma_{ge}\gamma_{bc}}),
\end{equation}
where $\gamma_{all}=\gamma_{ge}+\gamma_{eg}+\gamma_{bc}$. Approximated by a Taylor series, the ionization rate is then written as:
\begin{equation}\label{ionrate1}
  \gamma_{ion}\approx\frac{\gamma_{ge}\gamma_{bc}}{\gamma_{ge}+\gamma_{eg}+\gamma_{bc}}.
\end{equation}
As $\gamma_{ge}$ and $\gamma_{bc}$ both linearly depend on the laser power, the ionization rate would show a quardratical power dependence in the weak excitation region ($\gamma_{ge}+\gamma_{bc} \ll \gamma_{eg}$), and a linear power dependence in the strong excitation region ($\gamma_{ge}+\gamma_{bc} \gtrsim \gamma_{eg}$). For recombination process, the transitions are similar with those in ionization process. The form of recombination rate should be same with ionization rate.

\end{document}